\documentstyle[12pt,a4wide]{article}
\input{epsf}

\begin{document}
\title{Yang--Mills sphalerons in all even spacetime dimensions
$d=2k$, $k>2$ : $k$=3,4}
\author{{\large Yves Brihaye}$^{\ddagger}$
and {\large D. H. Tchrakian}$^{\dagger \star}$ \\ \\
$^{\ddagger}${\small Physique-Math\'ematique, Universite de 
Mons-Hainaut, Mons, Belgium}\\ \\
$^{\dagger}${\small Department of
Mathematical Physics, National University of Ireland Maynooth,} \\
{\small Maynooth, Ireland} \\
$^{\star}${\small School of Theoretical Physics -- DIAS, 10 Burlington
Road, Dublin 4, Ireland }}

\date{}
\newcommand{\dd}{\mbox{d}}
\newcommand{\tr}{\mbox{tr}}
\newcommand{\la}{\lambda}
\newcommand{\ka}{\kappa}
\newcommand{\al}{\alpha}
\newcommand{\ga}{\gamma}
\newcommand{\de}{\delta}
\newcommand{\si}{\sigma}
\newcommand{\bomega}{\mbox{\boldmath $\omega$}}
\newcommand{\bsi}{\mbox{\boldmath $\sigma$}}
\newcommand{\bchi}{\mbox{\boldmath $\chi$}}
\newcommand{\bal}{\mbox{\boldmath $\alpha$}}
\newcommand{\bpsi}{\mbox{\boldmath $\psi$}}
\newcommand{\brho}{\mbox{\boldmath $\varrho$}}
\newcommand{\beps}{\mbox{\boldmath $\varepsilon$}}
\newcommand{\bxi}{\mbox{\boldmath $\xi$}}
\newcommand{\bbeta}{\mbox{\boldmath $\beta$}}
\newcommand{\ee}{\end{equation}}
\newcommand{\eea}{\end{eqnarray}}
\newcommand{\be}{\begin{equation}}
\newcommand{\bea}{\begin{eqnarray}}
\newcommand{\ii}{\mbox{i}}
\newcommand{\e}{\mbox{e}}
\newcommand{\pa}{\partial}
\newcommand{\Om}{\Omega}
\newcommand{\vep}{\varepsilon}
\newcommand{\bfph}{{\bf \phi}}
\newcommand{\lm}{\lambda}
\def\theequation{\arabic{equation}}
\renewcommand{\thefootnote}{\fnsymbol{footnote}}
\newcommand{\re}[1]{(\ref{#1})}
\newcommand{\R}{{\rm I \hspace{-0.52ex} R}}
\newcommand{\N}{{\sf N\hspace*{-1.0ex}\rule{0.15ex}%
{1.3ex}\hspace*{1.0ex}}}
\newcommand{\Q}{{\sf Q\hspace*{-1.1ex}\rule{0.15ex}%
{1.5ex}\hspace*{1.1ex}}}
\newcommand{\C}{{\sf C\hspace*{-0.9ex}\rule{0.15ex}%
{1.3ex}\hspace*{0.9ex}}}
\newcommand{\eins}{1\hspace{-0.56ex}{\rm I}}
\renewcommand{\thefootnote}{\arabic{footnote}}

\maketitle

%\ \ \ PACS Numbers: 04.50.+h, 11.10.Kk, 11.15.Kc

\bigskip

\begin{abstract}
The classical solutions to higher dimensional Yang--Mills (YM) systems,
which form part of higher dimensional Einstein--YM (EYM) systems,
are studied. These are the gravity decoupling limits of the fully
gravitating EYM solutions. In odd spacetime dimensions, depending on the
choice of gauge group and its representation, these are either
topologically stable or are unstable.
Both cases are analysed, the latter numerically only.
In even spacetime dimensions they are always unstable, describing saddle
points of the energy, and can be described as {\it sphalerons}. This
instability is analysed by constructing the noncontractible loops and
calculating the Chern--Simons (CS) charges, and also perturbatively by
numerically constructing the negative modes. This study is restricted
to the simplest YM system in spacetime dimensions $d=6,7,8$, which
captures the qualitative features of the generic case.  
\end{abstract}
\medskip
\medskip
\newpage

\section{Introduction}
Gravitational and non Abelian gauge fields occur in low energy effective
actions~\cite{GSW} of superstring theory and supergravities. 
On the other hand classical solutions to the Einstein--Yang-Mills (EYM)
system, especially black holes, have an important role to play in
quantum gravity~\cite{HS}. These effective actions consist, in addition to
the usual Einstein-Hilbert and Yang--Mills (YM) systems, of
higher order terms in both the gravitational curvature, and the YM
curvature and its gauge covariant derivatives.
Moreover some of the theories in which such terms are
present are defined in higher dimensional spacetimes, namely on
$D$-branes~\cite{Pol}. Thus, the study of classical solutions to higher
curvature EYM models in higher dimensions is of physical interest.

It turns out that in higher (than $3+1$) dimensions, the higher order YM
curvature terms play a much more important role than do the corresponding
gravitational terms, e.g. Gauss-Bonnet terms. Inclusion of the latter does
not seem to alter the qualitative properties of the classical solutions,
while the absence of higher order YM terms prevents the existence of such
solutions due to the Derrick scaling requirements not being satisfied.
It is for this reason that we restrict our
considerations in this paper to the inclusion of higher order
YM~\footnote{Higher order gauge field curvatures arise also as quantum
corrections, see e.g. \cite{Polyak}, but this is not the source of
higher order terms we have in mind here.} only. Our
aim being the study of the stability (and its absence) in these systems,
we ignore the gravitational terms entirely since the most important
mechanism affecting stability/instability is characterised by the YM
sector alone.

Concerning the higher order YM curvature terms in the string theory
effective action, the situation is complex and as yet unresolved. While
YM terms up to $F^4$ arise from (the non Abelian version of) the
Born--Infeld action~\cite{Tseytlin}, it appears that this approach
does not yield all the $F^6$ terms~\cite{BRS}.
Terms of order $F^6$ and higher can also be obtained by employing the
constraints of (maximal) supersymmetry~\cite{CNT}. The results
of the various approaches are not identical.

Given the evolving stage which higher order curvature YM terms are in,
and motivated by the technical requirements for the construction of
classical solutions, we have restricted our considerations to one
particular family of higher curvature YM systems. The criterion is
that only the {\it second } power (and no higher power) of the velocity
field $\pa_0A_i$ occur in the Lagrangian. This constrains our choice
to one where the coefficients of each $F^{2n}$ term is fixed by the
requirement that only totally antisymmetrised curvature $n$-forms are
employed. We have no physical justification for this, but we hope that
the ensuing qualitative results hold also in the more general, and as yet
not definitely fixed, cases. In this family of YM systems the number of
higher order terms that can arise in any given dimension is limited
due to the imposed antisymmetry.

The other constraint we apply is that in any given dimension,
we truncate the series of $F^{2n}$ terms, to the minimum required to
satisfy the Derrick scaling requirement. Our justification here is that
higher order terms become more important at high energies so in the
low energy effective action it is sufficient to keep the lowest order
terms.

It is well known that there exist static regular~\cite{BK} and black
hole\cite{black1,black2} solutions to the Einstein--Yang-Mills (EYM)
system in $d=4$ spacetime dimensions.
To construct static solutions to gravitating Yang--Mills (YM) systems in
spacetime dimensions $d\ge 5$~\cite{bct2,bcht}, terms of higher order in
the YM curvature must be included. Such terms appear in various low
energy effective actions of string theory. For our practical purposes,
a particulary useful family of such YM systems is
\be
\label{YMhier}
{\cal L}_P=\sum_{p=1}^{P}\frac{\tau_p}{2(2p)!}\ \mbox{Tr}\,F(2p)^2
\ee
which is the sum of the terms in the YM hierarchy~\cite{T}, with $F(2p)$
being the $p$ fold totally antisymmetrised products of the YM curvature,
$F(2)=F_{\mu\nu}$, in this notation. Clearly, the highest value $P$
of $p$ in \re{YMhier} is finite and depends on the dimensionality $d$ of
the spacetime. In \cite{bct2,bcht} we chose the simplest possibility
$P=2$, since we restricted our study to $8\ge d\ge 5$. The study of the
classical solutions to the systems \re{YMhier}, which turn out to be the
gravity--decoupling solutions of EYM systems in these dimensions, is the
aim of this paper. This is an important aspect of the study of higher
dimensional EYM solutions.

To complete the definition of the models \re{YMhier} the gauge group $G$
must be specified. With the aim of constructing static spherically
symmetric solutions in $d$ spacetime dimensions, the smallest
such gauge group is $G=SO(d-1)$~\footnote{Should it turn out that the
gauge group must be $SO(n)$, $n<D$, it would only be possible to construct
solutions subject to less than
spherical symmetry.}. Static finite energy
solutions to the systems \re{YMhier} may or may not be topologically
stable depending on the dimensionality $d$ of the spacetime and the
choice of gauge group $G$.

The purpose of the present work is to resolve the question of stability
or instability of these spherically symmetric solutions quantitatively.
These are the gravity decoupling limits of the static and regular
solutions to EYM systems in spacetime dimensions $d\ge 6$~\cite{bct2}. In
the $d=5$ case~\cite{bcht}, as in the usual $d=4$ case~\cite{BK}, there
are no gravity decoupling solutions, unlike in the case of the EYM-Higgs
(EYMH) systems in $d=4$ whose regular solutions~\cite{bfm,w} tend to the
't~Hooft-Polyakov monopole in the flat space limit. In both cases,
higher dimensional EYM and EYMH in $d=4$, the models feature an
additional dimensional constant. As a result gravitating solutions exist
only for a finite range of values of the gravitational constant. Here, we
restrict to the non--gravitating models only.

For simplicity, we restrict to the model considered in \cite{bct2},
\re{YMhier} with $p=1$ and $2$ only, in spacetime dimensions $d=6,7$ and
$8$. Solutions to this model on $d=6$ Euclidean space were constructed in
\cite{but} in a different context. Consistently with our requirement for
the imposition of spherical symmetry we choose $G=SO(d-1)$ and $G=SO(d)$,
and because of the higher order of nonlinearity in the $p>1$ models the
choice of the concrete representations of $G$ is important.

Except for the case where there exists a nonvanishing Chern--Pontryagin
(CP) charge in the spacelike dimensions, these solutions are unstable. In
the particular cases where $G$ and its representation is chosen such that
the Chern--Simons (CS) charge is nonzero, this instability will be that
of a {\it sphaleron}~\cite{M,KM}. The quantitative study~\cite{AKY} of
the latter will be the major task below.

In section {\bf 2}, we present the detailed models and the corresponding
CS densities. In section {\bf 3}, spherical symmetry in the $d-1$
spacelike dimensions is imposed and the residual one dimensional models
are displayed in {\bf 3.1} where the gauge invariance of the Ansatz is
also stated. The static equations are given in {\bf 3.2} and the
corresponding CS densities in {\bf 3.3}. In {\bf 3.4}, the CS charges
are calculated. Section {\bf 4} contains our numerical results. In
{\bf 4.1} we have constructed the noncontractible loop (NCL) for the
unstable solutions and have plotted
the energy of the solution versus the angle parametrising this loop. In
those cases where there is a nonvanishing CS charge, the energy is plotted
also against the CS number. In {\bf 4.2} we have constructed the negative
modes excited by the instability and have calculated the negative
eigenvalues. Section {\bf 5} is devoted to a summary and discussion of
our results.

\section{The models and Chern-Simons densities}
In the first subsection {\bf 2.1} we define the static energy density
functional resulting from the Lagrangian \re{YMhier} in spacetime
dimensions $d=6,7,8$, and consider the topological lower bound on the
energy for the $d=7$ case. In the second subsection {\bf 2.2} we define
the Chern-Simons densities for the models in spacetimes $d=6$ and $8$.

Spacetime coordinates are labeled by Greek indices
$\mu ,\nu ,...$, and spacelike coordinates by Latin indices
$i,j,k,...$.

\subsection{The models}
Since we are interested in static solutions, we will define the
models on the $d-1$ dimensional Euclidean spacelike manifold. In the
present work, we will restrict our considerations to the $P=2$ model
\re{YMhier} used in \cite{bct2}. Expressed in component form this is
\be
\label{model}
{\cal H}_2=-\frac{\tau_1}{2.2!}\,\mbox{Tr}\,F_{ij}^2\,+\,
\frac{\tau_2}{2.4!}\,\mbox{Tr}\,F_{ijkl}^2
\ee
where
\[
F_{ijkl}\,=\,\{F_{i[j},F_{kl]}\}
\]
is the $4$-form curvature, and $[jkl]$ implies cyclic symmetrisation.
In the light of Derrick's scaling requirement, \re{model} is the simplest
model that can support static finite energy solutions for spacetime
dimensions up to $d=8$. Beyond that terms with higher values $p$ must be
included in \re{YMhier}, but this is entirely unnecessary since all our
conclusions from the present investigation hold qualitatively also for
$d\ge 9$.

The model is specified finally by the choice of the gauge group $G$, as
well as its representaion. We will mostly restrict ourselves to
$G=SO(d)$, but for {\it odd} $d$ we will include the special case
of $G=SO(d-1)$ as well.

When $d$ is even, i.e. $d=6$ and $8$ in our examples, there exists a
Chern--Simons (CS) density in the $5$ and $7$ spacelike dimensions and the
static solutions will display a sphaleron like unstability. As will be
seen below, the nonvanishing of the CS density is what necessitates the
choice of $G=SO(d)$, with $d=6$ and $8$. These
two examples will be the main focus of our attention.

When $d$ is odd, i.e. $d=7$ in our examples, there exists a
Chern-Pontryagin (CP) density in the $6$ spacelike dimensions. Provided
that $G$ and its representation is chosen suitably, this CP density is
nonvanishing and the resulting static solution will be stable. As will be
seen below, the existence of a stable soltion is what necessitates the
choice of $G=SO(6)(=SO(d-1))$ here. Otherwise, with $G=SO(7)(=SO(d))$, the
static solution will not be stable.

Before proceeding to state the CP and the CS densities for the even $d$
systems, we give the topological lower bound on the energy density
functional of the odd $d$, namely the $d=7$ system. The inequality
\be
\label{ineq}
\mbox{Tr}\left(\sqrt{\tau_1}F_{ij}-\frac{\sqrt{\tau_2}}{4!}
\vep_{ijklmn}F_{klmn}\right)^2\ge 0
\ee
leads to
\be
\label{lb}
{\cal H}_2\ge\frac18\sqrt{\tau_1\tau_2}\,\vep_{ijklmn}\,
\mbox{Tr}\,F_{ij}F_{kl}F_{mn}\,,
\ee
stating the topological lower bound in terms of the 3rd CP desity.
It follows that the spherically symmetric solution~\cite{but} to the model
in $d=7$ is topologically stable provided that the CP density on the
right hand side does not vanish. This is the case for $G=SO(d-1)=SO(6)$,
but only if the gauge fields are in the {\it chiral} represenation of $G$.
These static solutions in even spacelike dimensions, which obey
instanton like boundary conditions, play the role of solitons.
The
topological inequality \re{ineq} cannot be saturated since the
corresponding Bogomol'nyi equations are overdetermined.

\subsection{The Chern-Pontryagin and Chern-Simons densities}
The definition of Chern-Simons (CS) densities in $d-1$ spacelike
dimensions follows from that of the Chern-Pontryagin (CP) densities, in
{\it even} spacetime dimensions,
\be
\label{CP}
\varrho_d=\frac{1}{N_d}\vep_{\mu_1\mu_2\mu_3...\mu_{d}}
\mbox{Tr}\,F_{\mu_1\mu_2}F_{\mu_2\mu_3}...F_{\mu_{d-1}\mu_d}\quad
,\quad d=6\quad{\rm and}\quad d=8\,,
\ee
where $N_d$ is the appropriate normalisation factor for dimension $d$.
Of course, for the purpose of definition \re{CP}, the signature is
taken to be Euclidean.

The CP density is a total divergence, which we denote generally as
\be
\label{totdiv}
\varrho_d=\pa_{\mu}\,\Omega_{\mu}^{(d)}\,.
\ee
In the two cases of interest at hand, with $d=6$ and $d=8$,
$\Omega_{\mu}^{(d)}$ is given by
\bea
\Omega_{\mu_1}^{(6)}&=&\frac{5}{2^6\cdot\pi^3}
\vep_{\mu_1\mu_2\mu_3...\mu_{6}}\mbox{Tr}\,
A_{\mu_2}\left[F_{\mu_3\mu_4}F_{\mu_5\mu_6}-
F_{\mu_3\mu_4}A_{\mu_5}A_{\mu_6}+
\frac25A_{\mu_3}A_{\mu_4}A_{\mu_5}A_{\mu_6}\right]\label{CP6}\\
\Omega_{\mu_1}^{(8)}&=&\frac{7}{3\cdot 2^{12}\cdot\pi^4}
\vep_{\mu_1\mu_2\mu_3...\mu_{8}}\mbox{Tr}\,
A_{\mu_2}\bigg[F_{\mu_3\mu_4}F_{\mu_5\mu_6}F_{\mu_7\mu_8}
-\frac45F_{\mu_3\mu_4}F_{\mu_5\mu_6}A_{\mu_7}A_{\mu_8}-\frac25
F_{\mu_3\mu_4}A_{\mu_5}F_{\mu_6\mu_7}A_{\mu_8}\nonumber\\
&&\qquad\qquad\qquad\qquad\qquad\qquad
+\frac45F_{\mu_3\mu_4}A_{\mu_5}A_{\mu_6}A_{\mu_7}A_{\mu_8}-\frac{8}{35}
A_{\mu_3}A_{\mu_4}A_{\mu_5}A_{\mu_6}A_{\mu_7}A_{\mu_8}\bigg]\,.\label{CP8}
\eea
From \re{CP6} and \re{CP8}, the respective CS densities
$\nu_{d-1}$ are defined to be the $\mu=0$ components of
$\Omega_{\mu}^{(d)}$ in $5$ and $7$ Euclidean spacelike dimensions
\bea
\nu_5&=&-\frac{5}{2^6\cdot\pi^3}\vep_{i_1 i_2 i_3...i_5}\mbox{Tr}\,
A_{i_1}\left[F_{i_2 i_3}F_{i_4 i_5}-
F_{i_2 i_3}A_{i_4}A_{i_5}+
\frac25A_{i_2}A_{i_3}A_{i_4}A_{i_5}\right]\label{CS5}\\
\nu_7&=&-\frac{7}{3\cdot 2^{12}\cdot\pi^4}\vep_{i_1 i_2 i_3...i_7}
\mbox{Tr}\,A_{i_1}\bigg[F_{i_2 i_3}F_{i_4 i_5}F_{i_6 i_7}
-\frac45F_{i_2 i_3}F_{i_4 i_5}A_{i_6}A_{i_7}-\frac25
F_{i_2 i_3}A_{i_4}F_{i_5 i_6}A_{i_7}\nonumber\\
&&\qquad\qquad\qquad\qquad\qquad\qquad
+\frac45F_{i_2 i_3}A_{i_4}A_{i_5}A_{i_6}A_{i_7}-\frac{8}{35}
A_{i_2}A_{i_3}A_{i_4}A_{i_5}A_{i_6}A_{i_7}\bigg]\,.\label{CS7}
\eea

\section{Imposition of spherical symmetry}
Our choice of gauge group will be $G=SO(d)$, except in $d=7$ where we will
consider and dispose of the special case $G=SO(6)=SO(d-1)$. Our choice
of $G=SO(d)$ for even $d$ is similar to the choice for the
Weinberg--Salam~\cite{M} (WS) and for the Bartnik--McKinnon~\cite{VG1,V}
sphalerons in $d=4$. Furthermore, like in the latter case, we will employ
the {\it chiral} representation $G=SO_{\pm}(d)$, noting that for the
special case $d=4$, $G=SO_{\pm}(4)=SU_{L/R}(2)$.

Since our main aim is to study the instability of the solutions to
\re{model}, we will employ the general spherically
symmetric Ansatz parametrised by three radial functions $w_1(r)$,
$w_2(r)$ and $w_3(r)$. This follows from the axially symmetric
Ansatz~\cite{W} in $d$ dimensions where all the components of the gauge
connection are taken to be independent of $x_0$ and the component
$A_0=0$. Like the sphalerons~\cite{M,V} in $d=4$, our
solutions are also parametrised by
the function $w_1$ while the functions $(w_2,w_3)$ will be excited
only in the directions of the instability.

\medskip
\noindent
\underline{For even $d$ with $G=SO(d)$}, including $d=4$, the imposition
of spherical symmetry on the gauge field on $(d-1)$ dimensional Euclidean
space results in
\be
\label{sph.d.sig}
A_i=\frac{1-w_1(r)}{r}\ \Sigma_{ij}^{(d)}\,\hat x_j+
\frac{w_2(r)}{r}\,(\delta_{ij}-\hat x_i\hat x_j)\,\Sigma_{j,d}^{(d)}+
\frac{w_3(r)}{r}\,\hat x_i\hat x_j\,\Sigma_{j,d}^{(d)}
\ee
where $\Sigma_{ij}^{(d)}$ is (one of) the chiral representaion(s) of the
algebra of $SO(d)$, namely
\bea
\Sigma_{ij}^{(d)}&=&
\left(\frac{\eins\pm\Gamma_{d+1}}{2}\right)\Gamma_{ij}\label{sigma}\\
\Gamma_{ij}^{(d)}&=&-\frac14[\Gamma_i,\Gamma_j]\,,\label{gamma}
\eea
$\Gamma_{ij}$ being the spinor represention matrices of the algebra of
$SO(d)$ defined in terms of $\Gamma_i$, the gamma matrices in $d$
dimensions.

\medskip
\noindent
\underline{For odd $d$ with $G=SO(d)$}, there are no chiral
representations, so the Ansatz \re{sph.d.sig} holds only in a formal way,
replacing the matrices $\Sigma_{ij}^{(d)}$ by $\Gamma_{ij}^{(d)}$.

\medskip
\noindent
\underline{For odd $d$ with $G=SO(d)$}, it is possible to employ
the chiral representation of $SO(d-1)$. Here again, \re{sph.d.sig} holds
formally, but now by replacing $d$ with $d-1$ in it. As
noted above, in this case the solution (with $w_2=w_3=0$ everywhere) will
describe a stable soliton.

For the evaluation of the residual one dimensional energy density
functional, only the algebraic properties of $\Sigma_{ij}^{(d)}$
and of $\Gamma_{ij}^{(d)}$ enter the calculations, so the same result
holds (up to an unimportant numerical factor $2$) both for even and for
odd $d$ and with  all $G$. For the evaluation of the CP and CS densities
however, this distinction must be kept.

\subsection{Reduced one dimensional systems and gauge freedom}
Imposition of spherical symmetry on the $(d-1)$ dimensional $SO(d)$ system
results in the one dimensional energy density functional
\bea
H_d&=&
\frac{2^{\frac{d-2}{2}}\tau_1}{2\cdot2!}\,\frac{d-2}{4}\,r^{d-4}\cdot
\left[\left(w_1'+\frac{w_2w_3}{r}\right)^2+
\left(w_2'-\frac{w_1w_3}{r}\right)^2+
\frac{d-3}{2r^2}(1-\vert\vec w\vert^2)^2\right]\nonumber\\
&+&\frac{2^{\frac{d-2}{2}}\tau_2}{2\cdot4!}\,3^2\,(d-2)(d-3)(d-4)\,r^{d-8}
(1-\vert\vec w\vert^2)^2\cdot\nonumber\\
&&\qquad\qquad\cdot
\left[\left(w_1'+\frac{w_2w_3}{r}\right)^2+
\left(w_2'-\frac{w_1w_3}{r}\right)^2+
\frac{d-5}{4r^2}(1-\vert\vec w\vert^2)^2\right]\,,\label{onedimenergy}
\eea
in which we have used the shorthand notation
\[
\vert\vec w\vert^2=w_1^2+w_2^2\,.
\]

For odd $d=7$ and with $G=SO(6)$, \re{onedimenergy} holds as well, but the
solutions to the field equations (to be presented in the following
subsection) are quite different, with $w_1=w_2=0$ everywhere.

In the generic case, for the fields \re{sph.d.sig}, the requirement of
analyticity at the origin $r=0$ results in the asymptotic conditions
\be
\label{asym0}
w_1(0)=1\quad,\quad w_2(0)=0\quad,\quad w_3(0)=0\,,
\ee
while in the asymptotic region $r\gg 1$ the requirement of finiteness of
the energy results in the boundary condition
\be
\label{finitesod}
\lim_{r\rightarrow\infty}\vert\vec w\vert^2=
\lim_{r\rightarrow\infty}(w_1^2+w_2^2)^2=1\,.
\ee
\re{finitesod} can be parametrised, like in \cite{AKY}, in terms of the
familiar angle $q$ as
\be
\label{q}
\lim_{r\rightarrow\infty}w_1=\cos q\quad ,\quad
\lim_{r\rightarrow\infty}w_2=\sin q\,.
\ee
The conditions \re{finitesod}-\re{q} can be understood better by
displaying the gauge freedom of the energy density functional
\re{onedimenergy}, under the action of the local $U(1)$ transformation
\be
\label{u1}
g[\omega(r)]\in U(1)\quad ,\quad \lim_{r\to 0}\omega(r)=0\quad ,\quad
\lim_{r\to\infty}\omega(r)=-q\,.
\ee
The action of this $U(1)$ gauge transformation is given by
\bea
\left(
\begin{array}{c}
w_1\\
w_2
\end{array}
\right)&\stackrel{g[\omega]}\mapsto&
\left(
\begin{array}{cc}
\cos\omega & -\sin\omega \\
\sin\omega & \cos\omega
\end{array} \right)
\left(
\begin{array}{c}
w_1\\
w_2
\end{array}
\right)\label{trans12}\\
w_3&\stackrel{g[\omega]}\mapsto&w_3+r\,\omega'\label{trans3}\,.
\eea

%with
%\be
%\label{asymq}
%\lim_{r\to 0}q=0\quad ,\quad \lim_{r\to\infty}q=\pi\,.
%\ee
%\newpage
\subsection{The Euler--Lagrange equations}

The Euler-Lagrange equations of the $SO(d)$ systems are, for variations
of $w_1$, $w_2$ and $w_3$ in that order,
\bea
&\tau_1&\left\{ w_1''+\frac1r[(d-4)w_1'+2w_2'w_3+w_2w_3']+\frac{1}{r^2}
\left([(d-3)(1-\vert\vec w\vert^2)-w_3^2]w_1+(d-5)w_2w_3\right)\right\}
\nonumber\\
&+&\ \frac{3\tau_2}{4r^4}(d-3)(d-4)\ (1-\vert\vec w\vert^2)\cdot
\nonumber\\&&
\bigg\{(1-\vert\vec w\vert^2)
\left( w_1''+\frac1r[(d-8)w_1'+2w_2'w_3+w_2w_3']+\frac{1}{r^2}
\left([(d-5)(1-\vert\vec w\vert^2)-w_3^2]w_1+(d-9)w_2w_3\right)\right)
\nonumber\\&+&
2\left[\left(w_1'+\frac{w_2w_3}{r}\right)^2
+\left(w_2'-\frac{w_1w_3}{r}\right)^2\right]w_1
-2\left(\vert\vec w\vert^2\right)'\left(w_1'+\frac{w_2w_3}{r}\right)
\bigg\}=0\label{w1}
\eea
\bea
&\tau_1&\left\{ w_2''+\frac1r[(d-4)w_2'-2w_1'w_3-w_1w_3']+\frac{1}{r^2}
\left([(d-3)(1-\vert\vec w\vert^2)-w_3^2]w_2-(d-5)w_1w_3\right)\right\}
\nonumber\\
&+&\ \frac{3\tau_2}{4r^4}(d-3)(d-4)\ (1-\vert\vec w\vert^2)\cdot
\nonumber\\&\cdot&
\bigg\{(1-\vert\vec w\vert^2)\left( w_2''+
\frac1r[(d-8)w_2'-2w_1'w_3-w_1w_3']+\frac{1}{r^2}
\left([(d-5)(1-\vert\vec w\vert^2)-w_3^2]w_2-(d-9)w_1w_3\right)\right)
\nonumber\\&+&
2\left[\left(w_1'+\frac{w_2w_3}{r}\right)^2
+\left(w_2'-\frac{w_1w_3}{r}\right)^2\right]w_2
-2\left(\vert\vec w\vert^2\right)'\left(w_2'-\frac{w_1w_3}{r}\right)
\bigg\}=0\label{w2}
\eea

\medskip
\be
\label{w3}
\left[\tau_1+\frac{3\tau_2}{4r^4}(d-3)(d-4)(1-\vert\vec w\vert^2)^2\right]
\left[(w_1w_2'-w_2w_1')-\frac1rw_3\vert\vec w\vert^2\right]=0\ .
\qquad\qquad\qquad
\ee
The last equation, \re{w3}, is satisfied by the solutions of \re{w1} and
\re{w2}, as can be seen straighforwardly by differentiating \re{w3} and
identifying it with the difference of $w_2$ times \re{w1} and $w_1$ times
\re{w2}. Thus, exploiting the gauge freedom \re{trans12}-\re{trans3},
we choose the radial gauge with $w_3=0$. We then proceed to
solve the two equations \re{w1} and \re{w2} for the two functions $w_1$
and $w_2$.

We note here that in the $w_3=0$ gauge, equations \re{w1} and \re{w2}
are symmetric in the functions $w_1$ and $w_2$ as in the case of the
Bartnik-McKinnon sphaleron~\cite{VG1,V}, and in contrast to the
WS case~\cite{AKY} in which this symmetry is
absent due to the presence of the complex doublet Higgs field.

\subsubsection{$SO_{\pm}(d-1)$ solitons in $d$ dimensions}
Before proceeding to consider the sphaleron solutions, we dispose of
the stable soliton solutions of the models in {\it odd} $d$ spacetime
dimensions when the gauge connection takes its values in the {\it chiral}
representation of $SO(d-1)$. We restrict this demonstration to the
model in $d=7$, with $G=SO_{\pm}(6)$.

As a practical illustration of the topological lower bound \re{lb}, using
the notation $\ka_1=\sqrt{(d-2)\tau_1/2}$,
$\ka_2=3\sqrt{(d-2)(d-3)(d-4)\tau_2}$, consider the two inequalities
\bea
\left[\ka_1r^{\frac{d-4}{2}}w'
-\ka_2r^{\frac{d-8}{2}}\sqrt{\frac{d-5}{4r^2}}(1-w^2)^2\right]^2
&\ge&0\label{1>}\\
\left[\ka_2r^{\frac{d-8}{2}}(1-w^2)w'
-\ka_1r^{\frac{d-4}{2}}\sqrt{\frac{d-3}{2r^2}}(1-w^2)\right]^2
&\ge&0\label{2>}\ .
\eea
Adding \re{1>} and \re{2>} we have
\be
\label{totderiv}
H_d[w_1,w_2,w_3]=H_d[w,0,0]\ge 2\ka_1\ka_2\left(\sqrt{\frac{d-5}{4}}
+\sqrt{\frac{d-3}{2}}\right)r^{d-7}\,(1-w^2)\,w'\ ,
\ee
where $H_d[w,0,0]$ is given by the energy density functional
\re{onedimenergy} with $w_1=w$ and $w_2=w_3=0$. It is now obvious that
the right hand side of the inequality \re{totderiv} is a total derivative
only for the spacetime dimension $d=7$ at hand, and that it is the
residual CP density after the imposition of spherical symmetry on the
$G=SO_{\pm}(d-1)=SO_{\pm}(6)$ model, with $G$ is in the chiral
representation.

\subsection{Reduced Chern-Simons densities and non--contractible loop}
The reduced one dimensional Chern-Simons (CS) densities of static gauge
fields \re{sph.d.sig} in (even) $d$
spacetime dimensions, with gauge groups $G=SO(d)$ in the chiral
representation $SO_{\pm}(d)$, take on nonvanishing values~\footnote{Note
that if the representations of $G$ adopted were not the chiral ones given
by the spin matrices \re{sigma}, but rather those given by \re{gamma},
then the resulting CS densities would vanish.}. The
particular examples considered concretely here are those in $d=6$ and
$d=8$, given by \re{CS5} and \re{CS7}. Subjecting these to spherical
symmetry according to the Ansatz \re{sph.d.sig}, we find
\bea
n_5&=&-\frac{5}{2\pi}\left\{[2(1-w_1)+
3(1-\vert\vec w\vert^2)][(w_1-1)w_2'-w_2w_1']+
\frac{3}{r}\,w_3\,(1-\vert\vec w\vert^2)^2\right\}
\label{ncs5}\\
n_7&=&-\frac{7}{5\cdot 2^3\pi}\bigg\{
\left[5((1-w_1)^2+w_2^2)^2+6(1-w_1)(3(1-w_1)+4(1-|\vec w|^2)]\right]
[(w_1-1)w_2'-w_2w_1']\nonumber\\
&&\qquad\qquad\qquad\qquad\qquad\qquad\qquad\qquad\qquad\qquad\qquad\ \ \,
+\frac{5w_3}{r}(1-|\vec w|^2)^3
\bigg\}\,.
\label{ncs7}
\eea

In what follows a most important role will be played by the
non--contractible loop (NCL) of configurations displaying the instability
of the sphalerons. In contrast to the case of the WS model
\cite{AKY}, and similar to the Barnik--McKinnon sphaleron~\cite{VG1,V}.
This is a result of the $(w_1,w_2)$ symmetry in the field equations
\re{w1}-\re{w2}. In the former case where the doublet Higgs
field removes this $(w_1,w_2)$ symmetry, {\it conditional} solutions with
fixed CS number $Q_{CS}$ can be constructed by solving the equations of
motion that extremise the energy density functional plus a Lagrange
multiplier times the CS density. In the that case~\cite{AKY}, the
minimal finite energy path versus $Q_{CS}$ can be constructed concretely.
The situation here is similar, rather, to the second case of the
EYM sphaleron~\cite{VG1,V}, where the analysis of instability is carried
out exclusively by employing a NCL.

The actual sphaleron solutions are parametrised by the functions
\be
\label{sph}
w_1=f(r)\quad,\quad w_2=0\quad,\quad w_3=0
\ee
which will be constructed numerically in the next section,
consistently with the asymptotic conditions \re{asym0} and \re{finitesod},
the function $f(r)$ in \re{sph} satisfies the asymptotics
\be
\label{asymf}
\lim_{r\to 0}f(r)=1\quad,\quad\lim_{r\to\infty}f(r)=-1.
\ee
Following \cite{klink} we adopt the NCL configurations
\bea
\bar w_1&=&\frac12(1+\cos q)+\frac12(1-\cos q)\,f\label{NCLw1}\\
\bar w_2&=&\frac12(\sin q) \,(1-f)\label{NCLw2}\\
\bar w_3&=&0\,,\label{NCLw3}
\eea
which by virtue of \re{asymf} ensures that both \re{asym0} and \re{q}
are satisfied all along the NCL.

\subsection{Calculation of the Chern--Simons number}
In the case of the WS model~\cite{M,KM,AKY,V,VG1}, the CS number $Q_{CS}$
has the physical interpretation of Baryon number. Here, it merely is a
convenient topological charge characterising the sphaleron.

The topological CP charges are are given by the $d$ dimensional volume
integrals of $\pa_{\mu}\Omega_{\mu}^{(d)}$. Here we are interested in
the $k=3$ and $k=4$ examples for which $\Omega_{\mu}^{(d)}$ are given by
\re{CP6} and by \re{CP8} respectively, but now we consider static fields
in a $d$ dimensional Minkowskian spacetime. The topologial charge in this
context is referred to~\cite{KM} as the CS charge $Q_{CS}$.

Adapting the arguments of \cite{KM} to our cases, notably assuming that
$Q_{CS}=0$ at $t=-\infty$, the topological charges are given by
\be
\label{topch}
Q_{CS}=\frac{1}{N(d)}\left(\int_{-\infty}^{t_0} dt
\int_{S_{(d-1)}}\Omega_i^{(d)}\ dS_i+\int_{t=t_0}d^{d-1}x\ \Omega_0^{(d)}
\right)\,.
\ee
The Chern--Simons (CS) densities
$\Omega_0^{(d)}=\nu_{d-1}$ are displayed by \re{CS5}-\re{CS7}. The
task at hand is to evaluate the integrals \re{topch}
for our two sphaleron solutions.

In the $w_3=0$ gauge in which the sphaleron solutions are
constructed, the surface integral term in \re{topch} does not vanish
since the gauge potential decays with power $r^{-1}$ as $r\to\infty$,
as seen from \re{sph.d.sig} and  \re{sph}-\re{asymf}. It is convenient to
evaluate \re{topch} in a gauge in which the surface integral vanishes,
such that
\be
\label{topch-g}
Q_{CS}=\frac{1}{N(d)}\int_{t=t_0}d^{d-1}x\ \Omega_0^{(d)}
=\frac{1}{N(d)}\int_{t=t_0}\,\nu_{d-1}\,d^{d-1}x\,,
\ee
which for the spherically symmetric fields \re{sph.d.sig} reduces to the
one dimensional integral
\be
\label{topch1dim}
Q_{CS}=\int_{r=0}^{r=\infty}\,n_{d-1}\,dr\,
\ee
with $n_{d-1}(r)$ given by \re{ncs5}-\re{ncs7}.

But substituting the NCL configuration \re{NCLw1}-\re{NCLw3}
in the expressions for CS densities \re{ncs5} and \re{ncs7}, results
in the vanishing of $n_5$ and of $n_7$. This means that the nonvanishing
contribution to the integral \re{topch} must come from the surface
integral term, which is not convenient to evaluate since the solutions
and the NCL configurations at our disposal are time independent. But by
definition \re{topch} is gauge invariant, so it should be
evaluated in a gauge in which the surface integral in \re{topch} does not
contribute. To this end, following \cite{KM}, we subject
\re{NCLw1}-\re{NCLw3} to the gauge transformation
\re{trans12}-\re{trans3}, such that
\be
\label{o=-q}
\lim_{r\to 0}\omega(r)=0\quad,\quad\lim_{r\to\infty}\omega(r)=-q\ .
\ee
With these boundary conditions on the $U(1)$ gauge group parameter, and
\re{asymf} for the sphaleron profile function $f(r)$, we find  
\be
\label{sph-g}
\lim_{r\to\infty}{}^{\omega}\bar{w_1}=1\quad,\quad
\lim_{r\to\infty}{}^{\omega}\bar{w_2}=0\quad,\quad
\lim_{r\to\infty}{}^{\omega}\bar{w_3}=0\,,
\ee
which results in the connection \re{sph.d.sig} decaying faster than
$r^{-1}$ at infinity. As desired the surface integral term in \re{topch}
now vanishes. The density $n_{d-1}$ in \re{topch1dim} does not
vanish, and can be evaluated to
give the CS number.
%In the $d=4$ case we have
%\be
%\label{d4}
%Q_{CS}=\frac{1}{2\pi}\int_0^{\infty}\,dr\,\frac{d}{dr}
%\left[\omega-\frac12(1+f)\sin\omega-\frac12(1-f)\sin(q+\omega)
%-\frac12 f\sin q\right]
%\ee
%which by virtue of \re{u1} yields
%\be
%\label{cs4}
%Q_{CS}=\frac{q-\sin 2q}{2\pi}\,.
%\ee

Subjecting the NCL configuration
$(\bar{w_1},\bar{w_2},\bar{w_3})$ given by \re{NCLw1}-\re{NCLw3} to the
gauge transforamtion \re{trans12}-\re{trans3} with \re{o=-q}, and
substituting the resulting set of functions
$(^{\omega}\bar{w_1},^{\omega}\bar{w_2},^{\omega}\bar{w_3})$ into the
densities \re{ncs5} and \re{ncs7}, we obtain total derivatives in the
variable $r$. After integration, we have the respective CS
numbers~\footnote{In $d=4$,
$n_3=\frac{1}{2\pi}\left[(w_1-1)w_2'-w_2w_1'+\frac{w_3}{r}(1-|\vec w|^2)
\right]$, yielding the familiar~\cite{AKY} result
$Q_{CS}=\frac{q-\sin q}{2\pi}$.}
\bea
Q_{CS}&=&\frac{15}{2\pi}\left(3q - 4\sin q +
\frac12\sin 2q\right)\label{q5}\\
Q_{CS}&=&\frac{21}{20\pi}\left(30q - 45\sin q + 9\sin 2q -
\sin 3q\right)\label{q7}
\eea

\section{Numerical results}

\subsection{The sphaleron on the NCL}

We have solved the system  \re{w1}-\re{w2} numerically for $d=6,7,8$,
i.e. for the gauge group  $G=SO(d)$. (This excludes the case
of odd $d$ with $G=SO(d-1)$ with topological lower bound
\re{totderiv}).
The coupling constants $\tau_1, \tau_2$  can be chosen
arbitrarily by choosing an appropriate scale of the mass
$M$, defined as the integral of $H_d$
\[
M= \int H_d\,dr\ ,
\]
and of the radial variable $r$ in
(\ref{onedimenergy}). We have chosen $\tau_1=\tau_2 = 1$.
However, for practical reasons, 
the numerical values for the masses and for the negative
modes given below will be given in units $2^{(d-2)/2} S_{d-2}$,
where $S_{d-2}$ denotes the surface of the sphere in
the d-dimensional space-time.

The three profiles for the function $w_1(r)$ are 
presented in Figure 1, and  the figure reveals that
the dependence of the profile on $d$ is rather weak.
\begin{figure}
\epsfysize=22cm
\epsffile{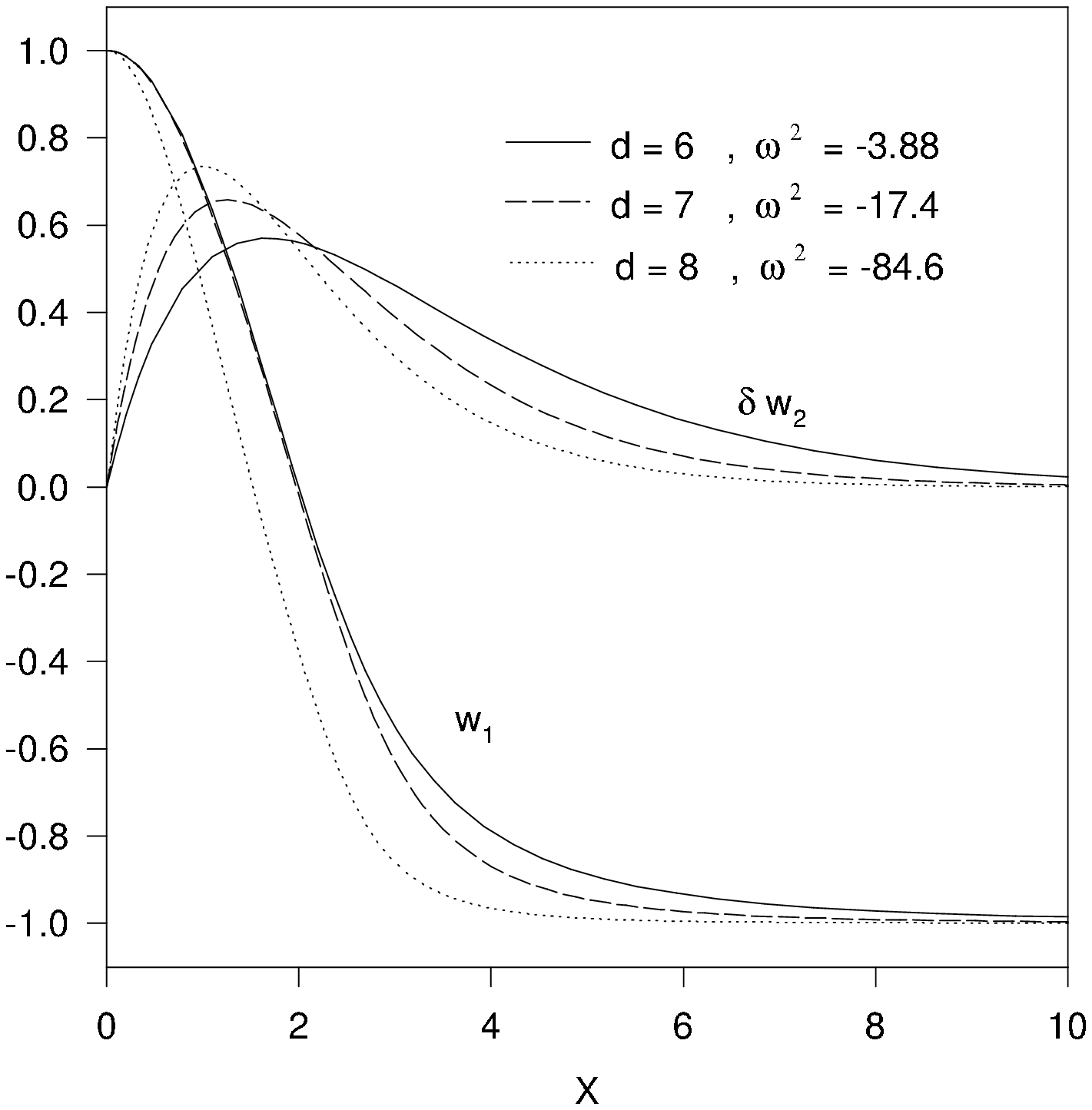}\vskip -1cm
\caption{\label{Fig.1} The profiles of the function $w_1$ for the
sphaleron solutions in $d=6,7,8$, which are practically unchanged when
the function $w_2$ is perturbed by $\delta w_2$. The profiles of the
negative modes $\delta w_2$ are displayed and the corresponding 
negative eigenvalues $\omega^2$ listed.}
\end{figure}

In the units choosen, the masses of these solutions are respectively
$M(d=6)\approx 4.67 $,
$M(d=7)\approx 11.3$,
$M(d=8)\approx 36.8$.
From the numerical profile for $w_1$,
the different configurations of the path
(\ref{NCLw1})-(\ref{NCLw3}) can be constructed;
their energies  can further be computed as functions of the parameter $q$.
The energy is plotted as a function of the parameter $q$
on Figure 2 for $d=6,7,8$.
\begin{figure}
\epsfysize=22cm
\epsffile{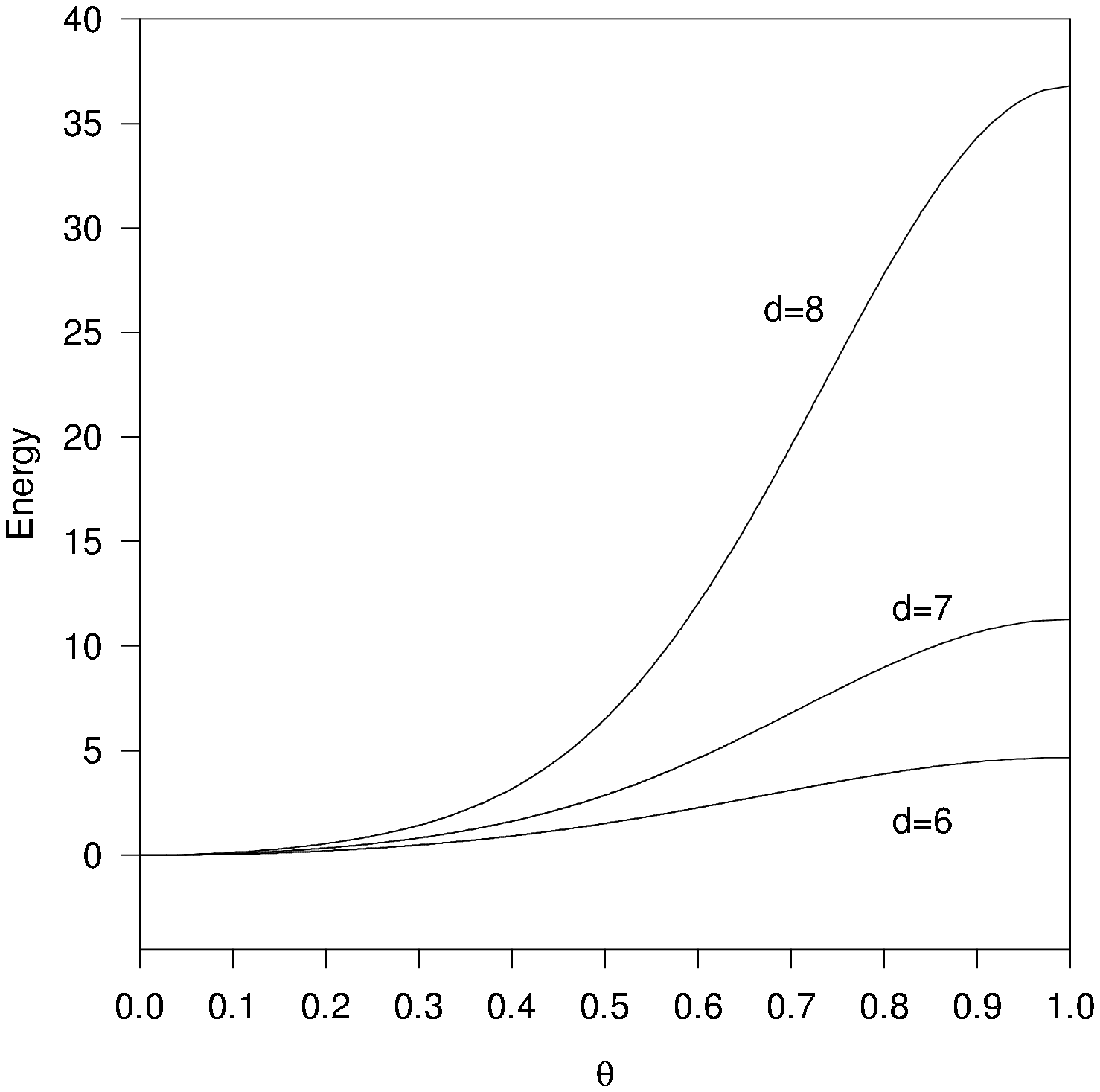}\vskip -1cm
\caption{\label{Fig.2} The energy along the path extrapolating
between the vacuum and the sphaleron is plotted as a function of
$\theta \equiv q/ \pi$ for d=6,7,8.
}
\end{figure}
As expected, the figure shows that the configurations on the
path have finite energy and that
the energy increases monotonically from the vacuum ($q=0$)
to the classical solution ($q = \pi$),
demonstrating that the classical solution is indeed a sphaleron.

For the cases with even $d$ and nonvanishing CS densities
\re{ncs5}-\re{ncs7}, the energy is plotted also against the CS charges
\re{q5}-\re{q7} respectively, on Figure 3.
\begin{figure}
\epsfysize=22cm
\epsffile{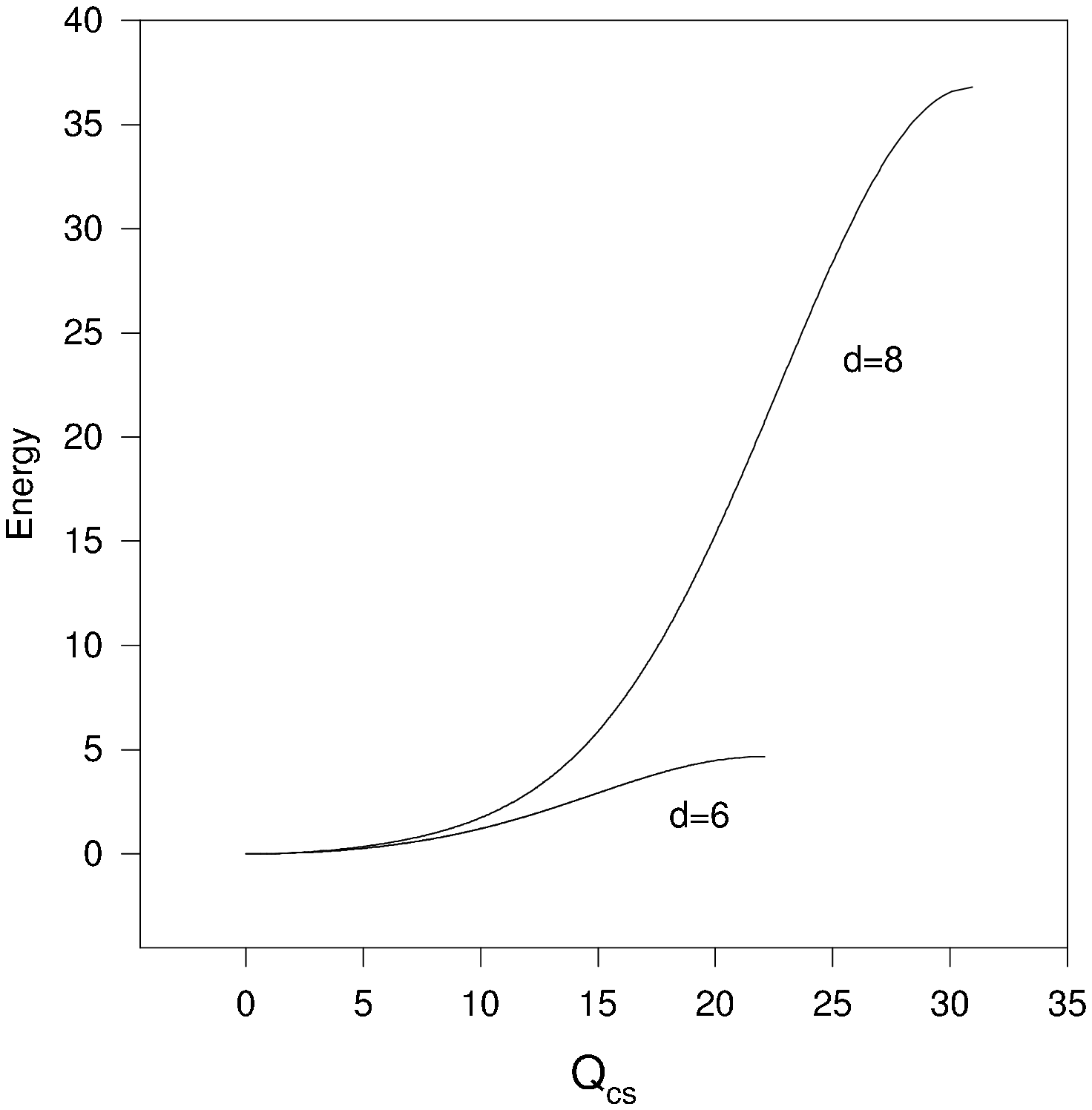}\vskip -1cm
\caption{\label{Fig.3} The energy along the path extrapolating
between the vacuum and the sphaleron is plotted as a function of 
$q$ for d=6,8.}
\end{figure}

\subsection{Negative modes}
In the second subsection we have carried out an infinitesimal stability
analysis by constructing the negative modes and evaluating their
negative eigenvalues. We also track, within the limits of validity of this
infinitesimal analysis, the growth of the CS charge as a function of
the $w_1$ and the function $w_2$ excited along the direction of
instability. For this purpose we consider a perturbation of the classical
solution constructed in the previous subsection~:
\be
\label{pertubate}
    \tilde  w_1 = w_1 + \e^{i \omega t} \delta w_1 \quad , \quad
    \tilde    w_2 = w_2 + \e^{i \omega t} \delta w_2  \quad , \quad
     \tilde   w_3=0\ ,
\ee
choosing again to work in the gauge $w_3=0$.
Inserting the perturbed solution into the equations of motion
and retaining only the linear terms in the fluctuation, we got
a system of two decoupled Sturm-Liouville  equations in
$\delta w_1$ and $\delta w_2$.
The construction of the normal modes is equivalent to
finding the normalisable solutions of these equations, which is a
problem beyond the scope of this paper. We limit
our analysis to the construction of the mode of lowest
eigenvalue. The results of \cite{yvesjutta} strongly
suggest that the main mode of instability of our solution
should  appear in the sector $\delta w_1=0$. This turns out
to be the case. Technically the negative mode can be constructed
by minimising the quadratic form
\be
    \delta E = 2^{(d-2)/2} S_{d-2} \frac{(d-2)}{16} \int dr r^{d-4} \left(
      I_1 (\delta w_2')^2 - I_2 (\delta w_2)^2
    \right)
\ee
 with
 \be
 \label{1}
 I_1 =  (\tau_1 +
       3 \tau_2 (d-3)(d-4)\frac{1}{r^4} (1-w_1^2)^2)
 \ee
 \be
 \label{2}
 I_2 =  (d-3) \frac{(1-w_1^2)}{r^2}
    \left( \tau_1    +
       3 \tau_2 (d-4) \frac{1}{r^4}
       ((d-5) (1-w_1^2)^2 + 2 r^2(w_1')^2)
             \right)\ .
\ee
In \re{1} -\re{2}, the function $w_1$ represents the classical
sphaleron solution. Inspecting the form of this variational problem we see
that it leads to a Sturm-Liouville equation
with a potential given by the function $-I_2$.
This corresponds to a potential well and allowing the existence
of a negative mode. We were able to construct numerically
one normalised negative mode $\delta w_2$ in each case $d=6,7,8$,
whose profiles are displayed in Figure 1, together with the profiles of
the corresponding classical solution $w_1$. The fact that $\delta w_2$
presents no node strongly suggests that our solution corresponds to the
eigenmode of lowest eigenvalue. These eigenvalues, indeed, appear to be
negative and were evaluated to be $-\omega^2 = 3.88, 17.4 , 84.6 $
respectively for d=6,7,8, again using the same scale as before.

Finally we evaluate the CS charge $Q_{CS}$ for $d=6$ case, for the
configurations of the form
$\tilde w_1 = w_1$ , $\tilde w_2 = \epsilon \delta w_2$,
$\epsilon$ being an infinitesimal parameter. We find
\be
    Q_{CS}(\tilde w_1 , \tilde w_2) =  Q_{CS}(q=\pi) - 1.54
    \epsilon
\ee
where $Q_{CS}(q=\pi)$ is the CS charge
of the sphaleron, which for $d=6$ is evaluated from \re{q5},
as $Q_{CS}(q=\pi)=22.5$. This shows that the CS charge varies linearly
with the  parameter $\epsilon$ when the sphaleron
is perturbed in the direction of its unstable mode.

\section{Summary and Conclusions} We have studied static finite energy
solutions to Yang--Mills systems in higher dimensions. The interest in these
solutions is that they are gravity decoupling limits of the fully
gravitating Einstein--Yang-Mills solutions in higher dimensions. In turn,
gravitating YM systems in higher dimensnions are important field theoretic
models arising in the study of $Dp$-branes in spacetime dimensions larger
than $4$, in the context of the low energy effective action of superstring
theory and gauged supergravities. Some of these solutions have sphaleron
type instabilities, which
must be studied. In this respect the higher dimensional gravitating YM
systems are similar to the gravitating YM model in $4$ spacetime dimensions,
whose sphaleron like instabilities were studied in detail in \cite{VG1,V}.
What is different in the higher dimensional cases at hand, versus the
$4$ spacetime dimensional case~\cite{VG1,V}, is that unlike for the latter,
here we have gravity decoupling limits, and, the sphaleron nature of the
gravitating solutions is essentially identical to their flat space
counterparts. The sphaleron analysis being appreciably simpler in the flat
case, we have chosen to work with those models.

We have restricted our studies to that of spherically symmetric solutions
only. This has necessitated certain, rather limited, choices of the
gauge groups $G$. These gauge groups turn out to be $SO(d-1)$ and
$SO(d)$ in {\it odd} spacetime dimensons $d$, and $SO(d)$ only in
{\it even} $d$. The gauge connections take their values in the spinor
reprentations in terms of Dirac matrices, and whenever $G=SO(N)$ is
defined for even $N$, we have employed the {\it chiral} representations
of $SO(N)$. All solutions considered are evaluated numerically and
they satisfy the second order Euler--Lagrange equations rather than
first order selfduality equations, since the YM systems in question
are not scale invariant.

We have shown that in {\it odd} spacetime dimensions $d$, the
solutions to $SO(d-1)$ YM systems are topologically stable if the
representaion of $G$ employed is {\it chiral}, i.e. that they
are solitons which are stabilised by the Chern--Pontryagin (CP)
topological charge. These are the only models which support stable
solitons. Should the representations of the $SO(d-1)$ algebra  employed
not be taken to be the {\it chiral} ones, the solutions will be unstable.
They are unstable also for the choice of $G=SO(d)$.

In {\it even} spacetime dimensions, the solutions turn out to be always
unstable, due essentially to the absence of a CP topological charge in
the {\it odd} spacelike dimensions. For $G=SO(d)$, by employing the
{\it chiral} representations, we have calculated the Chern--Simons (CS)
charges, and have plotted the energy of the noncontractible loops versus
the CS charge $Q_{CS}$, highlighting the nature of these solutions as
types of {\it sphalerons}.

For all the unstable solutions, namely to the $SO(d)$ models in all
$d$ dimensional spacetimes, we have constructed the negative modes of the
corresponding fluctuation equations
and calculated their negative eigenvalues.
For the $SO_{\pm}(6)$ solutions in $d=6$ in particular, we have also
calculated $Q_{CS}$ perturbatively, showing that in the region of
validity the $Q_{CS}$ increases linearly in the pertubation parameter.

We have restricted our study to the simplest model \re{YMhier}, with
$p=1$ and $p=2$ terms only, in spacetime dimensions $d=6,7,8$. This
limited choice is sufficient to illustrate the questions of stability and
instability in the generic cases. The properties demonstrated in the $p=2$
examples studied repeat themselves in every $4p$ dimensions. Thus for
example for $p=3$ models, in spacetime dimensions $d=6,8,10,12$ the solutions
will be unstable sphalerons, while those in $d=7,9,11$ will be stable,
being stabilised by the 3rd, 4th and 5th Pontryagin charges, respectively.
If the $d=7$ model chosen in that case does not contain the $p=2$ (in addition
to the obligatory $p=3$ term dictated by Derrick)term in
the YM hierarchy, then there will be no gravity decoupling limit. In that case
the sphaleron analysis must be carried out in the fully gravitating model,
following the lines of the analysis in \cite{VG1,V} for $d=4$. The same holds
true in the case of the $p=2$ model studied here, $d=5$. In that case too there
exists no flat space limit so that the sphaleron analysis must again be carried
out as in $p=1$, $d=4$~\cite{V,VG1}. The latter analysis falls outside the
scope of the present work and is deferred to a future study.

\medskip
\medskip

\noindent
{\large\bf{Acknowledgements}}
We thank D. Maison for a discussion which gave rise to the present study,
and E. Radu for his generous cooperation in the course of this work.
This work is supported by Enterprise--Ireland Basic Science Research
project SC/2003/390. Y.B. wishes to thank the Belgian FNRS for financial
support.

\newpage

\begin{small}

\end{small}

\begin{thebibliography}{99}
\bibitem{GSW}
M.B. Green, J.H. Schwarz and E. Witten, Superstring Theory,
Cambridge University Press, Cambridge, 1987.

\bibitem{HS}
J. Harvey and A. Strominger, ``TASI lectures on quantum aspects of black
holes'', hep-th/9209055.

\bibitem{Pol}
J. Polchinski, ``TASI lectures on D-branes'', hep-th/9611050.

\bibitem{Polyak}
A.M. Polyakov, Nucl. Phys. B {\bf 120} (1977) 459.

\bibitem{Tseytlin}
A.A. Tseytlin, {\it Born--Infeld action, suersymmetry and string theory},
in Yuri Golfand memorial volume, ed. M. Shifman, World Scientific (2000).

\bibitem{BRS}
E. Bergshoeff, M. de Roo and A. Sevrin, Fortsch.Phys. 49 (2001) 433-440;
Nucl.Phys.Proc.Suppl. 102 (2001) 50-55.

\bibitem{CNT}
M. Cederwall, B. Nilsson and D. Tsimpis, JHEP 0106 (2001) 034.

\bibitem{BK}
R. Bartnik and J. McKinnon, Phys. Rev. Lett. {\bf 61} (1988) 141.

\bibitem{black1}
M.S. Volkov and D.V. Gal'tsov, JETP Lett. {\bf 50} (1989) 346.

\bibitem{black2}
P. Bizon, Phys. Rev. Lett. {\bf 64} (1990) 2844.

\bibitem{bct2}
Y. Brihaye, A. Chakrabarti and D.H. Tchrakian, Class. Quant. Grav.
{\bf 20} (2003) 2765 [hep-th/0202141]

\bibitem{bcht}
Yves Brihaye, A. Chakrabarti, Betti Hartmann and D.H. Tchrakian,
Phys. Lett. B {\bf 561} (2003) 161 [hep-th/0212288]

\bibitem{T}
see D.H. Tchrakian  Yang-Mills hierarchy, Int. J. Mod. Phys. A
(Proc.Suppl.) {\bf 3A} (1993) 584, and references therein.

\bibitem{bfm}
P. Breitenlohner, P. Forgacs and D. Maison, Nucl. Phys. B
{\bf 383} (1992) 357; {\it ibid.} {\bf 442} (1995) 126.

\bibitem{w} K. Lee, V. P. Nair and E. J. Weinberg, Phys. Rev. D {\bf 45}
(1992) 2751.

\bibitem{but}
J. Burzlaff and D. H. Tchrakian, J. Phys. A {\bf 26} (1993) L1053.

\bibitem{M}
N.S. Manton, Phys. Rev. D {\bf 28} (1983) 2019.

\bibitem{KM}
F.R. Klinkhamer and N.S. Manton, Phys. Rev. D {\bf 30} (1984) 2212.

\bibitem{AKY}
T. Akiba, H. Kikuchi and T. Yanagida, Phys. Rev. D {\bf 38} (1988) 1937.

\bibitem{W}
E. Witten, Phys. Rev. Lett. {\bf 38} (1978) 121.

\bibitem{VG1}
D.V. Gal'tsov and M.S. Volkov, Phys. Lett. B  {\bf273} (1991) 255.

\bibitem{V}
M.S. Volkov, Phys. Lett. B {\bf 334} (1994) 40.

%\bibitem{VG}
%M.S. Volkov and D.V. Gal'tsov, Phys. Rep. {\bf 319} (1999) 1;
%hep-th/9810070.

\bibitem{klink}
F. R. Klinkhamer, Phys. Lett. B {\bf 236} (1990) 187.

%%\bibitem{colsys}
%%U. Asher, J. Christiansen and R. D. Russel,
%%Math. Comput. 33 (1979) 659;
%%ACM Trans. Math. Softw. 7 (1981) 209.

\bibitem{yvesjutta}
Y. Brihaye and J. Kunz, Phys. Lett. B {\bf 249} (1990) 90.

\end{thebibliography}
\end{document}